\documentclass[twocolumn,showpacs,preprintnumbers,amsmath,amssymb]{revtex4}

\usepackage{graphicx}
\usepackage{dcolumn}
\usepackage{bm}
\usepackage{amssymb}
\usepackage{amsfonts}
\usepackage{amsmath}
\usepackage{latexsym}
\usepackage{dsfont}
\usepackage{multirow}
\usepackage{color}
\usepackage[mathscr]{eucal}

\definecolor{nv}{rgb}{0.1,0.1,0.6}
\definecolor{pr}{rgb}{0.2,0.1,0.5}
\definecolor{mg}{rgb}{0.4,0.0,0.4}

\newcommand{\beq}{\begin{equation}}
\newcommand{\eeq}{\end{equation}}
\newcommand{\beqy}{\begin{eqnarray}}
\newcommand{\eeqy}{\end{eqnarray}}
\newcommand{\beqyn}{\begin{eqnarray*}}
\newcommand{\eeqyn}{\end{eqnarray*}}

\newcommand{\bs}{\begin{slide}}
\newcommand{\es}{\end{slide}}
\newcommand{\bc}{\begin{center}}
\newcommand{\ec}{\end{center}}
\newcommand{\bmin}{\begin{minipage}}
\newcommand{\emin}{\end{minipage}}

\newcommand{\bi}{\begin{itemize}}
\newcommand{\ei}{\end{itemize}}

\newcommand{\bea}{\begin{eqnarray}}
\newcommand{\eea}{\end{eqnarray}}
\newcommand{\be}{\begin{equation}}
\newcommand{\ee}{\end{equation}}

\newcommand{\uTr}{\mathrm{Tr}}

\newcommand{\uslash}{/\!\!\!}

\newcommand{\uvec}[1]{\boldsymbol{#1}}

\begin{document}


\title{New explicit expressions for Dirac bilinears}

\author{C\'edric Lorc\'e}
\email{cedric.lorce@polytechnique.edu}
\affiliation{Centre de Physique Th\'eorique, \'Ecole polytechnique, 
	CNRS, Universit\'e Paris-Saclay, F-91128 Palaiseau, France}

\date{\today}

\begin{abstract}
We derive new explicit expressions for the Dirac bilinears based on a generic representation of the massive Dirac spinors with canonical polarization. These bilinears depend on a direction $n$ in Minkowski space which specifies the form of dynamics. We argue that such a dependence is unavoidable in a relativistic theory with spin, since it originates from Wigner rotation effects. Contrary to most of the expressions found in the literature, our ones are valid for all momenta and canonical polarizations of the spinors. As a by-product, we also obtain a generic explicit expression for the covariant spin vector.
\end{abstract}

\maketitle

\section{Introduction}

In particle physics, one often faces the problem of computing amplitudes which are expressed as
\begin{equation}\label{bilinear}
\mathcal M_\Gamma=\overline w_{\epsilon'}(p',S')\Gamma w_\epsilon(p,S),
\end{equation}
where $\Gamma$ is a $4\times 4$ matrix in Dirac space, $w_+(p,S)=u(p,S)$ and $w_-(p,S)=v(p,S)$ are the free positive and negative-energy spinors, and $p$ ($p'$) is the initial (final) four-momentum of the particle with mass $m$ ($m'$). The covariant polarization vectors $S$ and $S'$ satisfy the usual onshell relations
\begin{equation}\label{propS}
S^2=S'^2=-1,\qquad p\cdot S=p'\cdot S'=0.
\end{equation}

An old question is whether the amplitude $\mathcal M_\Gamma$ can be expressed as an analytic and covariant function depending on the vectors $p$, $p'$, $S$, and $S'$. The main motivation for this is that such a covariant expression greatly helps the calculations in practical cases and allows one to develop efficient numerical methods, see e.g.~\cite{Bondarev:1997kf} and references therein.

Recently, hadronic physics provided a new motivation. Indeed, there is a growing interest in spatial distributions of quarks and gluons inside the nucleon, see e.g.~\cite{Burkardt:2000za,Burkardt:2002ks,Polyakov:2002yz,Diehl:2005jf,Miller:2010nz,Lorce:2011kd,Lorce:2015sqe,Lorce:2017wkb}. Since these distributions are expressed as Fourier transforms of $\mathcal M_\Gamma$ with respect to the four-momentum transfer $\Delta=p'-p$, explicit expressions for the Dirac bilinears in momentum space appear to be mandatory. So far, one has essentially focused on particular components of Lorentz tensors and used the non-tensorial expressions derived e.g. in~\cite{Lepage:1980fj,Brodsky:1997de}. More tensorial expressions would represent a major step forward allowing one to push analytical calculations further while preserving manifest Lorentz covariance as much as possible.
\newline

The aim of this paper is to derive new explicit expressions for the 16 Dirac bilinears $\mathcal M_{\Gamma^A}$ with $\Gamma^A=\{\mathds 1,\gamma_5,\gamma^\mu,\gamma^\mu\gamma_5,i\sigma^{\mu\nu}\}$, particularly well suited for hadronic purposes. The plan of the paper is the following. In Sec.~\ref{sec2} we quickly review the main methods used to compute amplitudes of the type~\eqref{bilinear}. In Sec.~\ref{sec3} we discuss relations based on charge conjugation and \textsf{CPT} symmetry, and we present the complete set of onshell identities arising from the free Dirac equation. In Sec.~\ref{sec4}, we derive new expressions for the Dirac bilinears, and discuss in detail the question of Lorentz covariance in Sec.\ref{sec5}. We then conclude the paper with Sec.~\ref{sec6}.

\section{Some standard methods}\label{sec2}

Casimir's trick allows one to evaluate the amplitude~\eqref{bilinear} up to a global phase, by converting its square into a trace in Dirac space~\cite{Casimir:1933,Griffiths:1987}. In the generalized form which does not require summation over polarizations, one can write
\begin{equation}
|\mathcal M_\Gamma|^2=\uTr[(\uslash p'+\epsilon'm')\,\tfrac{\mathds 1+\gamma_5\uslash S'}{2}\,\Gamma\,(\uslash p+\epsilon m)\,\tfrac{\mathds 1+\gamma_5\uslash S}{2}\,\overline \Gamma]
\end{equation}
using the Dirac adjoint $\overline \Gamma=\gamma^0\Gamma^\dag\gamma^0$ and the projection operator onto $w_\epsilon$ spinors with definite momentum and polarization
\begin{equation}\label{projector}
w_\epsilon\overline w_\epsilon=(\uslash p+\epsilon m)\,\tfrac{\mathds 1+\gamma_5\uslash S}{2}.
\end{equation}
While this method is very general and provides manifestly covariant expressions, the evaluation of the trace becomes rapidly tedious in practice when $\Gamma$ is a product of a large number of Dirac matrices. Moreover, the information about the phase of the amplitude $\mathcal M_\Gamma$ is lost in the procedure, which is problematic e.g. when interference contributions need to be computed. For these reasons, it is preferable to use another method that does not require to square the original amplitude.

A strategy, developed about 50 years ago~\cite{Bellomo:1961,Bogush:1962,Bjorken:1966kh,Henry:1967jm,Fearing:1972pt,Krivoruchenko:1993iq,Bondarev:1993mw,Galynskii:1999fj}, consists in multiplying $\mathcal M_\Gamma$ by the unity in the form $1=\mathcal M^*_{\Gamma'}/\mathcal M^*_{\Gamma'}$, where $\mathcal M_{\Gamma'}$ is of the form~\eqref{bilinear} with an arbitrary matrix $\Gamma'$. One can then write
\begin{equation}\label{Fearing}
\mathcal M_\Gamma=\tfrac{1}{\mathcal M^*_{\Gamma'}}\uTr[(\uslash p'+\epsilon'm')\,\tfrac{\mathds 1+\gamma_5\uslash S'}{2}\,\Gamma\,(\uslash p+\epsilon m)\,\tfrac{\mathds 1+\gamma_5\uslash S}{2}\,\overline{\Gamma'}].
\end{equation}
Casimir's trick then appears as the special case $\Gamma'=\Gamma$. In practice, the phase of $\mathcal M_{\Gamma'}$ is often not known, so that the denominator is replaced by $|\mathcal M_{\Gamma'}|$. It then follows that different choices for $\Gamma'$ lead to covariant expressions differing by a phase~\cite{Bondarev:1997kf}.

One can simplify the calculation further. Since $\Gamma$ is a matrix in Dirac space, it can be decomposed onto the basis $\Gamma^A$
\begin{equation}\label{Diracdecomposition}
\Gamma=\tfrac{1}{4}\sum_{A=1}^{16}\uTr[\Gamma\Gamma_A]\,\Gamma^A,
\end{equation}
where $\Gamma_A$ is the dual basis satisfying $\tfrac{1}{4}\,\uTr[\Gamma^A\Gamma_B]=\delta^A_B$. The amplitude in Eq.~\eqref{bilinear} can then be written as a linear combination of the Dirac bilinears $\mathcal M_{\Gamma^A}$
\begin{equation}
\mathcal M_\Gamma=\tfrac{1}{4}\sum_{A=1}^{16}\uTr[\Gamma\Gamma_A]\,\mathcal M_{\Gamma^A}.
\end{equation}
Clearly, this method requires the evaluation of traces with a smaller number of Dirac matrices compared to Casimir's trick. The problem reduces to finding a covariant expression for the Dirac bilinears, which can be computed from Eq.~\eqref{Fearing} with $\Gamma=\Gamma^A$. Explicit expressions for $\Gamma'=\mathds 1$ can be found in~\cite{Fearing:1972pt}. For a review of the different choices of $\Gamma'$ in the literature, see~\cite{Bondarev:1997kf}. 

The main issue with this approach is that it does not provide a valid covariant expression for all vectors $p$, $p'$, $S$, and $S'$. Indeed, whenever $\mathcal M_{\Gamma'}$ vanishes, one obtains from Eq.~\eqref{Fearing} indeterminate results of the type $0/0$. One could of course use different expressions in different regions of $p$, $p'$, $S$, and $S'$ to avoid these indeterminate forms, but one then faces the problem of matching these expressions with each other because of unknown phase factors. Another option would be to use an explicit representation for the Dirac spinors and matrices, and to compute explicitly the Dirac bilinears, see e.g.~\cite{Lepage:1980fj,Brodsky:1997de,Hagiwara:1985yu}. The problem however is that it makes the calculations complicated with bulky and non-covariant expressions.

The failure of finding generic covariant expressions motivated the development of the two-component formalism, see~\cite{Dixon:1996wi,Dreiner:2008tw} for a review. The idea is to express the amplitude $\mathcal M_\Gamma$ in the massless case not in terms of Lorentz tensors, but in terms of spinor products like $\overline u(p',h')u(p,h)$ with helicities $h,h'=\pm$, considered as more fundamental objects. This approach is rather simple to implement and provides powerful compact expressions when fermion masses can be neglected. Generalizations to the massive case do exist, see e.g.~\cite{Kleiss:1985yh,Kleiss:1986ct,Hagiwara:1986vm,Ballestrero:1994jn,Andreev:2001se}, but usually turn out to be quite cumbersome in practice.

\section{General relations}\label{sec3}

Before we derive our new explicit expressions for the Dirac bilinears, we collect in this section a set of identities that can be used either to reduce the number of necessary expressions, or to test their consistency.  

We label the spinors $u\equiv u(p,\lambda)$ and $\overline u'\equiv\overline u(p',\lambda')$ with the rest-frame spin projection (along the $z$-direction as usual) index $\lambda=\pm$ instead of the covariant polarization vector $S$. The reason is that current hadronic applications, like e.g. the spatial distributions mentioned in the introduction, require a clear separation between spin and orbital angular momentum, which can only be achieved with respect to the rest frame.

For a spin-$1/2$ particle with rest-frame spin along the direction represented by the unit three-vector $\uvec s$, the covariant spin-density matrix is given by
\begin{equation}\label{spindens}
\overline u(p,\lambda')\gamma^\mu\gamma_5u(p,\lambda)=2m\,S^\mu_{\lambda'\lambda}(p)
\end{equation}
and is related to the covariant spin vector as follows
\begin{equation}
S^\mu(p,\uvec s)=\tfrac{1}{2}\sum_{\lambda',\lambda}(\uvec s\cdot\uvec\sigma)_{\lambda\lambda'}S^\mu_{\lambda'\lambda}(p),
\end{equation}
where $\uvec\sigma$ are the three Pauli matrices. The covariant spin vector provides the explicit map between a rest-frame polarization along $\uvec s$ and a covariant polarization vector in the frame where the particle has four-momentum $p$. The inverse relation can be put in the form
\begin{equation}\label{genericS}
S^\mu_{\lambda'\lambda}(p)=S^\mu(p,\uvec \sigma_{\lambda'\lambda}).
\end{equation}
In some sense, the covariant spin vector and the covariant spin-density matrix can be viewed as the same object, expressed in two different bases related by Pauli matrices: one labeled by the pair $\lambda',\lambda$ and one by the unit vector $\uvec s$. Note that this connection is made for pure states only.

\subsection{Discrete symmetries}

Using the standard relative phase conventions given in Appendix B of~\cite{Diehl:2003ny}, positive and negative-energy Dirac spinors are related by charge conjugation symmetry
\begin{equation}
w_\epsilon=C\,\overline w^T_{-\epsilon}
\end{equation}
with the antisymmetric matrix $C=-C^T$ satisfying $C^\dag=C^{-1}$ and
\begin{equation}
C\gamma^{\mu T} C^{-1}=-\gamma^\mu,\qquad C\gamma^T_5 C^{-1}=\gamma_5.
\end{equation}
We can also write owing to $\textsf{CPT}$ symmetry
\begin{equation}
w_\epsilon=-\epsilon\gamma_5\widetilde w_{-\epsilon},
\end{equation}
where the flipped spinors are defined as
\begin{equation}
\widetilde w_\epsilon(p,\lambda)=\lambda\,w_\epsilon(p,-\lambda),\qquad \lambda=\pm.
\end{equation}

Thanks to these relations, we can express all Dirac amplitudes in terms of $\overline u'\Gamma u$ only. Indeed, from charge conjugation symmetry, we get
\begin{align}
\overline v'\Gamma v&=-\overline u\, C\Gamma^TC^{-1}u',\\
\overline v'\Gamma u&=-\overline v \,C\Gamma^TC^{-1}u',\\
\overline u'\Gamma v&=-\overline u \,C\Gamma^TC^{-1}v',
\end{align}
and from $\textsf{CPT}$ symmetry we have
\begin{align}
\overline v'\Gamma v&=-(\overline{\tilde u}\,\!'\gamma_5\Gamma\gamma_5\tilde u),\label{vvCPT}\\
\overline u'\Gamma v&=\overline u'\Gamma\gamma_5\tilde u,\\
\overline v'\Gamma u&=-\overline{\tilde u}\,\!'\gamma_5\Gamma u.
\end{align}
Therefore, in the following it will be sufficient to restrict our considerations to the positive-energy sector.

\subsection{Onshell identities}

We now provide the complete list of onshell identities among the Dirac bilinears, which generalize the famous Gordon identity~\cite{Gordon:1928}.

Using the Dirac equation for positive-energy Dirac spinors $(\uslash p-m)u=0$, we find that for any matrix $\Gamma$ in Dirac space
\begin{align}
\overline u'\Gamma u&=\tfrac{1}{\bar m}\,\overline u'\left(\{\,\uslash\! P,\Gamma\}+\tfrac{1}{2}\,[\,\uslash\!\Delta,\Gamma]\right)u,\label{id1}\\
0&=\overline u'\left(\tfrac{1}{2}\,\{\,\uslash\! \Delta,\Gamma\}+[\,\uslash\!P,\Gamma]\right)u,\label{id2}
\end{align}
where $\bar m=(m'+m)/2$ and
\begin{equation}
P=\tfrac{\bar m}{2}\,\big(\tfrac{p'}{m'}+\tfrac{p}{m}\big),\qquad \Delta=\bar m\,\big(\tfrac{p'}{m'}-\tfrac{p}{m}\big),
\end{equation}
which satisfy $P\cdot\Delta=0$ and $P^2+\tfrac{\Delta^2}{4}=\bar m^2$. Inserting the (overcomplete) basis in Dirac space $\Gamma=\mathds 1,\gamma_5,\gamma^\mu,\gamma^\mu\gamma_5,i\sigma^{\mu\nu},i\sigma^{\mu\nu}\gamma_5$ in Eq.~\eqref{id1}, we obtain a whole set of onshell identities ($\epsilon_{0123}=+1$)
\begin{align}
\overline u'u&=\overline u'\tfrac{\,\uslash \!P}{\bar m}u,\label{Scalar}\\
\overline u'\gamma_5u&=\overline u'\tfrac{\,\uslash \!\Delta\gamma_5}{2\bar m}u,\label{Pseudo}\\
\overline u'\gamma^\mu u&=\overline u'\left[\tfrac{P^\mu}{\bar m}+\tfrac{i\sigma^{\mu\Delta}}{2\bar m}\right]u,\label{Vector}\\
\overline u'\gamma^\mu\gamma_5 u&=\overline u'\left[\tfrac{\Delta^\mu\gamma_5}{2\bar m}+\tfrac{i\sigma^{\mu P}\gamma_5}{\bar m}\right]u,\label{Axial}\\
\overline u'i\sigma^{\mu\nu} u&=\overline u'\left[-\tfrac{\Delta^{[\mu}\gamma^{\nu]}}{2\bar m}+\tfrac{i\epsilon^{\mu\nu P\tau}\gamma_\tau\gamma_5}{\bar m}\right]u,\label{Tensor}\\
\overline u'i\sigma^{\mu\nu}\gamma_5 u&=\overline u'\left[-\tfrac{P^{[\mu}\gamma^{\nu]}\gamma_5}{\bar m}+\tfrac{i\epsilon^{\mu\nu \Delta\tau}\gamma_\tau}{2\bar m}\right]u,\label{Tensor5}
\end{align}
where we used the convenient notations $i\sigma^{\mu a}=i\sigma^{\mu\rho}a_\rho$, $i\epsilon^{\mu\nu a\tau}=i\epsilon^{\mu\nu \rho\tau}a_\rho$, and $a^{[\mu}b^{\nu]}=a^\mu b^\nu-a^\nu b^\mu$. Because of the matrix identity
\begin{equation}
i\sigma^{\mu\nu}=-\tfrac{1}{2}\,i\epsilon^{\mu\nu\alpha\beta}i\sigma_{\alpha\beta}\gamma_5,
\end{equation}
the onshell identity~\eqref{Tensor5} is equivalent to~\eqref{Tensor}. We nevertheless included it in our list in order to emphasize the symmetry between bilinears that have opposite behavior under parity transformation. Indeed, one obtains one from the other by means of the substitutions $\mathds 1\leftrightarrow \gamma_5$ and $P \leftrightarrow \Delta/2$.

Inserting now the same basis in Eq.~\eqref{id2}, we get
\begin{align}
0&=\overline u'\tfrac{\,\uslash\!\Delta}{2} u,\label{S0}\\
0&=\overline u'\,\uslash\!P\gamma_5 u,\\
0&=\overline u'\left[\tfrac{\Delta^\mu}{2}+i\sigma^{\mu P}\right]u,\\
0&=\overline u'\left[P^\mu\gamma_5+\tfrac{i\sigma^{\mu \Delta}\gamma_5}{2}\right]u,\\
0&=\overline u'\left[-P^{[\mu}\gamma^{\nu]}+\tfrac{i\epsilon^{\mu\nu \Delta\tau}\gamma_\tau\gamma_5}{2}\right]u,\\
0&=\overline u'\left[-\tfrac{\Delta^{[\mu}\gamma^{\nu]}\gamma_5}{2}+i\epsilon^{\mu\nu P\tau}\gamma_\tau\right]u.\label{T50}
\end{align}
These onshell identities are not new as they can alternatively be derived from Eqs.~\eqref{Scalar}-\eqref{Tensor5} either by contraction or antisymmetric (exterior) product with $P^\mu$ and $\Delta^\mu$.

Note that in the negative-energy sector, the Dirac equation reads $(\uslash p+m)v=0$, implying that the onshell identities among $\overline v'\Gamma v$ are the same as among $\overline u'\Gamma u$ with the substitution $\bar m\leftrightarrow -\bar m$. This simple relation can also be understood from Eq.~\eqref{vvCPT} using dimensional analysis together with $\{\gamma^\mu,\gamma_5\}=0$.

We close this section with a side remark. In the general covariant parametrization of nucleon matrix elements, one usually makes use of the Gordon identity~\eqref{Vector} and its parity partner~\eqref{Axial} to eliminate the $\gamma^\mu$ and $\gamma^\mu\gamma_5$ Dirac structures, see e.g.~\cite{Meissner:2009ww}. We note that using the other onshell identities~\eqref{Scalar}, \eqref{Pseudo}, \eqref{Tensor} and \eqref{Tensor5}, there exists an alternative approach where one makes use of the Dirac structures $\gamma^\mu$ and $\gamma^\mu\gamma_5$ \emph{only}, as noted earlier by Yehudai~\cite{Yehudai:1991}.

\section{New explicit expressions}\label{sec4}

We now derive new explicit expressions for the Dirac bilinears. Our strategy consists in writing the amplitude~\eqref{bilinear} in the positive-energy sector as a trace
\begin{equation}\label{Trace}
\mathcal M_\Gamma=\uTr[\Gamma u\overline u'],
\end{equation}
together with a convenient representation of the outer product $u\overline u'$.

\subsection{Dirac spinors}

The massive positive-energy Dirac spinor with canonical polarization can conveniently be written as~\cite{Chung:1991st,deAraujo:1999ugw,Rinehimer:2009yv}
\begin{equation}\label{genericspinor}
u(p,\lambda)=\mathcal N_p\left(\,\uslash p+m\right)\uslash n\,\chi_\lambda,
\end{equation}
where $n$ is a timelike or a lightlike four-vector, and $\mathcal N_p$ is a normalization factor such that
\begin{equation}
\overline u(p,\lambda')u(p,\lambda)=2m\,\delta_{\lambda'\lambda}.
\end{equation}
We will implicitly work with the standard representation of the Dirac matrices, but our results easily generalize to any representation. The form~\eqref{genericspinor} can be obtained e.g. by boosting (without any rotation) the rest-frame spinors $\chi_\lambda$, which satisfy $\gamma^0\chi_\lambda=\chi_\lambda$ and $\overline\chi_{\lambda'}\chi_\lambda=\delta_{\lambda'\lambda}$, to the appropriate momentum $\uvec p$. Alternatively, Eq.~\eqref{genericspinor} can be seen as the projection of the reference spinor $\uslash n\,\chi_\lambda$ onto spinors with four-momentum $p$ and mass $m$ by means of the completeness relation
\begin{equation}
\sum_\lambda u(p,\lambda)\overline u(p,\lambda)=\uslash p+m.
\end{equation}

The key point we would like to stress here is that Lorentz boosts are necessarily used in the explicit construction of the Dirac spinors in momentum space with canonical polarization, see e.g.~\cite{Polyzou:2012ut}. The reason is because canonical polarization is defined with respect to a reference frame. Boosts are distinguished from rotations provided that the ``time'' variable $x\cdot n$, specified by the direction $n$, has already been defined. As shown by Dirac~\cite{Dirac:1949cp}, four-dimensional Minkowski space-time can be foliated into ``time'' and ``space'' in different ways that are not connected by Lorentz transformations. These different foliations are known as different \emph{forms} of dynamics. By making this dependence explicit, we can easily reduce our results to particular types of canonical Dirac spinors. For $n_\text{IF}=(1,0,0,0)$, we obtain the usual spinors in instant form (IF), and for $n_\text{LF}=(1,0,0,-1)/\alpha$ with $\alpha$ an arbitrary positive constant, we obtain the light-front (LF) spinors~\cite{Lepage:1980fj,Brodsky:1997de}. The sets of instant form and front form spinors constitute two different examples of free Dirac spinor basis with given momentum $p$. They can be expressed into each other through a so-called Melosh rotation~\cite{Melosh:1974cu,Ahluwalia:1993xa,Lorce:2011zta,Li:2015hew}
\begin{equation}
u_\text{LF}(p,\lambda)=\sum_\sigma u_\text{IF}(p,\sigma)D^M_{\sigma\lambda}(p),
\end{equation}
where $D^M_{\sigma\lambda}(p)=\overline u_\text{IF}(p,\sigma)u_\text{LF}(p,\lambda)/2m$ is a unitary matrix. A more detailed discussion about the interpolation between different types of canonical Dirac spinors can be found e.g. in~\cite{Li:2015hew}.

\subsection{Outer product}

The next step is to consider the outer product of two positive-energy Dirac spinors $u\overline u'$. Starting with the rest-frame spinors $\chi_\lambda$, we find the outer product
\begin{equation}
\chi\overline\chi'=\tfrac{\mathds 1+\gamma^0}{2}\,\tfrac{\mathds 1+\gamma_5\,\slash\!\!\!\!\Sigma}{2},
\end{equation}
where we omitted polarization indices and introduced $\Sigma^\mu=(0,\uvec\sigma)$. The outer product $\chi\overline \chi'$ can therefore be seen as the tensor product of a $4\times 4$ matrix in Dirac space with a $2\times 2$ matrix in polarization space. Summing over the polarizations amounts to considering the trace in polarization space, leading to $\text{tr}_2[\Sigma^\mu]=0$. Note the explicit appearance of the $0$-component which reflects the special role played by the rest frame. 

The generic form~\eqref{genericspinor} allows us to write down the outer product of two positive-energy Dirac spinors in terms of Dirac matrices
\begin{equation}\label{ubaru}
u\overline u'=\mathcal N_p\mathcal N_{p'}\left(\,\uslash p+m\right)\uslash n\,\tfrac{\mathds 1+\gamma^0}{2}\,\tfrac{\mathds 1+\gamma_5\,\slash\!\!\!\!\Sigma}{2}\,\uslash n\left(\,\uslash p'+m'\right).
\end{equation}
In the forward case $p'=p$ with equal masses $m'=m$, this should reduce to the Bouchiat-Michel expression~\cite{Bouchiat:1958}
\begin{equation}\label{bouchiatmichel}
u(p,\lambda)\overline u(p,\lambda')=\left(\uslash p+m\right)\tfrac{[\mathds 1\delta_{\lambda'\lambda}+\gamma_5\uslash S_{\lambda'\lambda}]}{2},
\end{equation}
which is simply the projection operator~\eqref{projector} expressed in terms of the rest-frame spin projection indices. Introducing for convenience the tensor
\begin{equation}
\Sigma_{\alpha\beta}=(\Sigma\cdot n)\,\delta^0_\alpha n_\beta-n^0\Sigma_\alpha n_\beta-\tfrac{n^2}{2}\,\delta^0_\alpha\Sigma_\beta,
\end{equation}
we obtain from the matching between Eqs.~\eqref{ubaru} and~\eqref{bouchiatmichel} the following general expression
\begin{equation}\label{fourpol}
S^\mu=\tfrac{2}{m}\,\Sigma_{\alpha\beta}\left[mp^{[\alpha}\eta^{\beta]\mu}+\eta^{\alpha0}(p^\beta p^\mu-\eta^{\beta\mu}m^2)\right]\mathcal N^2_p
\end{equation}
for the covariant spin-density matrix appearing in Eq.~\eqref{bouchiatmichel}, where $\eta_{\mu\nu}=\text{diag}(+1,-1,-1,-1)$ is the standard metric in Minkowski space, and
\begin{equation}
\mathcal N^2_p=\left[2 (p\cdot n)n^0+n^2(m-p^0)\right]^{-1}
\end{equation}
for the normalization factor introduced in Eq.~\eqref{genericspinor}. Note that the phase of $\mathcal N_p$ is not determined and has to be fixed by convention in practice.

It is clear that the covariant spin-density matrix is transverse onshell $p\cdot S=0$. Moreover, Eq.~\eqref{fourpol} reduces to the well-known expressions in instant form for $n=n_\text{IF}$
\begin{equation}\label{SIF}
S^\mu_\text{IF}(p)=\left(\tfrac{\uvec p\cdot\uvec\sigma}{m},\uvec\sigma+\tfrac{\uvec p\,(\uvec p\cdot\uvec\sigma)}{ m(E_p+ m)}\right),
\end{equation}
and in front form for $n=n_\text{LF}$
\begin{equation}\label{SLF}
S^\mu_\text{LF}(p)=\left[\tfrac{p^+}{m}\,\sigma_z,\tfrac{\uvec p_\perp^2-m^2}{2 mp^+}\,\sigma_z+\tfrac{\uvec p_\perp\cdot\uvec \sigma_\perp}{p^+},\uvec \sigma_\perp+\tfrac{\uvec p_\perp}{ m}\,\sigma_z\right],
\end{equation}
where we used the LF components $a^{\mu}=[a^+,a^-,\uvec a_\perp]$ with $a^-=a\cdot \bar n_\text{LF}$ and the dual lightlike direction $\bar n_\text{LF}=\alpha\,(1,0,0,1)/2$ which satisfies $n_\text{LF}\cdot\bar n_\text{LF}=1$. It follows from Eq.~\eqref{genericS} that the corresponding covariant spin vectors read explicitly
\begin{align}
S^\mu_\text{IF}(p,\uvec s)&=\left(\tfrac{\uvec p\cdot\uvec s}{m},\uvec s+\tfrac{\uvec p\,(\uvec p\cdot\uvec s)}{ m(E_p+ m)}\right),\\
S^\mu_\text{LF}(p,\uvec s)&=\left[\tfrac{p^+}{m}\,s_z,\tfrac{\uvec p_\perp^2-m^2}{2 mp^+}\,s_z+\tfrac{\uvec p_\perp\cdot\uvec s_\perp}{p^+},\uvec s_\perp+\tfrac{\uvec p_\perp}{ m}\,s_z\right].
\end{align}
Note that in Eq.~\eqref{ubaru}, we have considered that $u$ and $\overline u'$ are of the same type, i.e. they are defined with the same four-vector $n'=n$. One can of course consider the more general case with $n'\neq n$, but this reduces to the case $n'=n$ accompanied by a (generalized) Melosh rotation~\cite{Melosh:1974cu,Ahluwalia:1993xa,Lorce:2011zta,Li:2015hew}.

\subsection{Dirac bilinears}

We are now ready to derive the new explicit expressions for the Dirac bilinears. Using the representation~\eqref{ubaru} for the outer product in Eq.~\eqref{Trace} with $\Gamma=\mathds 1,\gamma_5,\gamma^\mu,\gamma^\mu\gamma_5,i\sigma^{\mu\nu},i\sigma^{\mu\nu}\gamma_5$, we find after some algebra
\begin{align}
\overline u'u&=2\left[\bar m\, (P\cdot N)+\Sigma_{\alpha\beta}\,i\epsilon^{\alpha\beta P\Delta}\right]\mathcal N,\label{S1}\\
\overline u'\gamma_5u&=2\,\Sigma_{\alpha\beta}\left(P^{0\alpha}\Delta^\beta-P^{\beta}\Delta^\alpha\right)\mathcal N,\\
\overline u'\gamma^\mu u&=2\left[Q^{\mu\tau}N_\tau-\Sigma_{\tau\beta} P^\tau_{\phantom{\tau}\alpha}\,i\epsilon^{\mu\alpha\beta\Delta}\right]\mathcal N,\\
\overline u'\gamma^\mu\gamma_5 u&=4\left[\tfrac{1}{4}\,i\epsilon^{\mu N P\Delta}+\Sigma_{0\tau}Q^{\tau\mu}\right.\nonumber\\
&\qquad\left.+\bar m\,\Sigma_{\alpha\beta}\left(P^{0\alpha}\eta^{\beta\mu}-\eta^{\mu\alpha}P^\beta\right)\right]\mathcal N,\\
\overline u'i\sigma^{\mu\nu}u&=4\left[\tfrac{\bar m}{4}\,N^{[\mu}\Delta^{\nu]}-\bar m\,\Sigma_{\tau\beta}P^\tau_{\phantom{\tau}\alpha}\,i\epsilon^{\mu\nu\alpha\beta}\right.\nonumber\\
&\quad\left.+\Sigma_{\alpha\beta}\left(i\epsilon^{\mu\nu\alpha\beta}\,\tfrac{\Delta^2}{4}+Q_\tau^{\phantom{\tau}[\mu}i\epsilon^{\nu]\alpha\beta\tau}\right)\right]\mathcal N,\\
\overline u'i\sigma^{\mu\nu}\gamma_5u&=4\left[-\tfrac{\bar m}{4}\,i\epsilon^{\mu\nu N\Delta}-\bar m\,\Sigma^{0[\mu}P^{\nu]}+\tfrac{\Delta^2}{4}\,\Sigma^{[\mu\nu]}\right.\nonumber\\
&\qquad\left.+\Sigma^{\tau[\mu}Q^{\nu]}_{\phantom{\nu]}\tau}+Q^{\tau[\mu}\Sigma^{\nu]}_{\phantom{\nu]}\tau}\right]\mathcal N,\label{T51}
\end{align}
where we defined for convenience the tensors
\begin{align}
N_{\alpha}&=2n^0 n_\alpha+\tfrac{n^2}{\bar m}\,P_{0\alpha},\\
P_{\alpha\beta}&=\delta_\alpha^0P_\beta-\bar m\eta_{\alpha\beta},\\
Q_{\alpha\beta}&=P_\alpha P_\beta+\tfrac{\Delta^2\eta_{\alpha\beta}-\Delta_\alpha\Delta_\beta}{4},
\end{align}
and the global normalization factor
\begin{equation}
\mathcal N=\mathcal N_{p'}\mathcal N_p\,\tfrac{m'm}{\bar m^2}.
\end{equation}
More explicit expressions in instant form ($n=n_\text{IF}$) and front form ($n=n_\text{LF}$) are given in the Appendix.

It is straightforward to check that our explicit expressions~\eqref{S1}-\eqref{T51} satisfy all the onshell relations~\eqref{Scalar}-\eqref{Tensor5} and \eqref{S0}-\eqref{T50}. Moreover, in the forward limit $p'=p$ with equal masses $m'=m$, they reduce to the well-known ones
\begin{align}
\overline u'u&=2m,\\
\overline u'\gamma_5u&=0,\\
\overline u'\gamma^\mu u&=2p^\mu,\\
\overline u'\gamma^\mu\gamma_5 u&=2m\,S^\mu,\label{axialSdef}\\
\overline u'i\sigma^{\mu\nu}u&=2m\,i\epsilon^{\mu\nu pS},\\
\overline u'i\sigma^{\mu\nu}\gamma_5u&=-2m\,p^{[\mu}S^{\nu]},
\end{align}
as expected.

\section{Discussion}\label{sec5}

Although we succeeded in obtaining explicit expressions for the Dirac bilinears in terms of Lorentz tensors, they do not look quite covariant at first sight, because they involve explicitly some $0$-component. Moreover, they also require some auxiliary four-vector $n$ unrelated to the kinematics of the process under study. 

We will argue in this section that, strictly speaking, our results are actually Lorentz covariant. Covariance is however not explicit due to the fact that our expressions are given in terms of the canonical polarizations.

\subsection{Lorentz covariance and tensors}

We use Lorentz covariance in its original meaning, namely as the property that equations take the same form in any Lorentz frame. In elementary treatments, textbooks on relativistic fields make use of this covariance property to determine the Lorentz transformation law of fields from the equations of motion. 

Covariance is particularly manifest if one succeeds in expressing all quantities in terms of Lorentz (or Dirac) tensors. Since this happens in so many cases, it seems that in practice covariance is usually identified with tensorial nature. Because of this identification, one occasionally runs however into conceptual conundrums. 

The typical example is electrodynamics in the Coulomb gauge, $\uvec\nabla\cdot\uvec A=0$, which is often presented as a non-covariant gauge. The theory has however been shown to lead to perfectly Lorentz covariant results~\cite{Bjorken:1965zz,Manoukian:1987hy}. To make sense of this, one sometimes rather says that electrodynamics in the Coulomb gauge is not \emph{explicitly} Lorentz covariant. All this boils down to the a priori Lorentz four-vector nature of the gauge potential $A_\mu$. Note however that because of gauge symmetry, the Lorentz transformation law of the gauge potential is only fixed \emph{up to a gauge transformation}~\cite{Weinberg:1995mt,Bjorken:1965zz,Manoukian:1987hy,Moriyasu:1984mh,Lorce:2012rr,Leader:2013jra}. If one uses the canonical formalism to derive from the theory itself the Lorentz transformation law of the gauge potential, one obtains indeed a vector potential that does not transform as a simple four-vector, but one which preserves the form of the gauge-fixing condition in the new frame, see e.g.~\cite{Bjorken:1965zz,Manoukian:1987hy} for the Coulomb gauge case. 

This is unfortunately not so well known because many standard textbooks gloss over these subtleties and simply (sometimes implicitly) assume that the gauge potential is a four-vector for pure convenience. Some popular textbooks, like e.g.~\cite{Feynman:1963uxa,Jackson:1998nia}, even present a proof of the four-vector nature of the gauge potential, but flaws jeopardizing this proof have been identified~\cite{Lorce:2012rr,Leader:2013jra}. 

Let us stress however that treating in practice the gauge potential as a four-vector is perfectly fine, since the non-tensorial part of the Lorentz transformation can always be compensated by a gauge transformation. In doing so, the advantage is that Lorentz covariance is preserved explicitly, but the price to pay is the introduction of non-physical degrees of freedom which must be constrained so as not to contribute to the physical observables.

The bottom line is that Lorentz covariance does not necessarily imply tensorial character. Probably the most blatant example is the connection $\Gamma^\alpha_{\phantom {\alpha}\mu\beta}$ in General Relativity which is not a tensor under general coordinate transformations, despite appearances. Explicit non-tensorial expressions for the Dirac bilinears have been derived in~\cite{Lepage:1980fj,Brodsky:1997de}. They are however perfectly Lorentz covariant since they were obtained from the standard spinors.

\subsection{Polarization basis}

Tensorial (i.e. explicitly covariant) expressions for the Dirac bilinears in terms of the Lorentz four-vectors $p$, $p'$, $S$ and $S'$ can be obtained using the standard methods summarized in Sec.~\ref{sec2}. These are however not adapted to the hadronic applications since the latter require a clear distinction between spin and orbital angular momentum, which is provided by a canonical basis. We therefore need an explicit connection between the covariant polarization vector $S^\mu$ and the rest-frame spin projection axis $\uvec s$, given by the covariant spin vector $S^\mu(p,\uvec s)$.

As a consequence of the non-commutativity of relativistic boosts, a canonical polarization basis can unambiguously be defined only in connection with some reference frame~\cite{Polyzou:2012ut,Wigner:1939cj,Weinberg:1995mt}. For a massive particle, the rest frame appears to be the most natural one. Once a polarization basis is fixed in the rest frame, one can unambiguously define a canonical polarization basis in any frame by a applying a standard boost (a conventional subset of the Lorentz group). For example, we have 
\begin{equation}
u(p,\lambda)\propto S[\Lambda_\text{st}(p)]\,\chi_\lambda,
\end{equation}
where $\Lambda_\text{st}(p)$ is the standard boost from $p^\mu_0=(m,\uvec 0)$ to $p^\mu$. Similarly, an explicit expression for the covariant spin vector can be obtained following the same procedure
\begin{equation}
S^\mu(p,\uvec s)=[\Lambda_\text{st}(p)]^\mu_{\phantom{\mu}\nu}S^\nu_0(\uvec s)
\end{equation}
with $S^\mu_0(\uvec s)=(0,\uvec s)$. Depending on the choice of the standard boost, one ends up with different types of canonical polarization, like e.g. canonical spin, helicity, or light-front helicity, see~\cite{Polyzou:2012ut,Wigner:1939cj,Weinberg:1995mt} for more details. The canonical polarization basis for a moving state therefore (implicitly) depends on the rest frame and on the definition of the standard boost.

Our explicit results are written within some canonical polarization basis, and so one should naturally expect some dependence on the rest frame and standard boosts to show up. Unlike~\cite{Lepage:1980fj,Brodsky:1997de}, we succeeded in obtaining expressions in terms of tensor which made these dependences \emph{explicit} in the form of $0$-components and auxiliary four-vector $n$. Note also that, as shown in the Appendix, our generic spinors reduce to the standard explicit spinors in both instant form and front form. This indicates that our results ought to be Lorentz covariant.

\subsection{Wigner rotations}

To understand better the Lorentz covariance of our results, let us consider the Lorentz transformation of a polarized state with four-momentum $p$. Any Lorentz transformation $\Lambda$ can be decomposed as follows
\begin{equation}
\Lambda=\Lambda_\text{st}(\Lambda p)R_W(\Lambda,p)\Lambda^{-1}_\text{st}(p),
\end{equation}
where the quantity
\begin{equation}\label{RW}
R_W(\Lambda,p)=\Lambda^{-1}_\text{st}(\Lambda p)\Lambda\Lambda_\text{st}(p)
\end{equation}
maps $p_0$ to $p_0$ and is known as the Wigner rotation. Since by definition the transformations $\Lambda_\text{st}(\Lambda p)$ and $\Lambda^{-1}_\text{st}(p)$ do not affect the polarization basis, the Lorentz transform of the basis spinor can be expressed as~\cite{Polyzou:2012ut}
\begin{equation}\label{Wignerrot}
\begin{aligned}
u'(p',\lambda)&=S[\Lambda]\,u(p,\lambda)\\
&=\sum_{\lambda'}u(\Lambda p,\lambda')\,D_{\lambda'\lambda},
\end{aligned}
\end{equation}
where $D_{\lambda'\lambda}$ represents the Wigner rotation matrix $R_W$ in two-dimensional polarization space. Note, in passing, that Vega and Wudka in~\cite{Vega:1995cc} were forced to include such a Wigner rotation, because they directly expressed $u(p',\lambda')$ in terms of $u(p,\lambda)$.

Using now Eq.~\eqref{spindens}, the Lorentz transform of the covariant spin-density matrix reads
\begin{equation}
\begin{aligned}
S'^\mu_{\lambda'\lambda}(p')&=\Lambda^\mu_{\phantom{\mu}\nu}S^\nu_{\lambda'\lambda}(p)\\
&=\sum_{\sigma',\sigma}D^\dag_{\lambda'\sigma'}S^\mu_{\sigma'\sigma}(\Lambda p)D_{\sigma\lambda}.
\end{aligned}
\end{equation}
The corresponding relation for the covariant spin vector is then
\begin{equation}
\begin{aligned}
S'^\mu(p',\uvec s)&=\Lambda^\mu_{\phantom{\mu}\nu}S^\nu(p,\uvec s)\\
&=S^\mu(\Lambda p,\mathcal R\uvec s),
\end{aligned}
\end{equation}
where $\mathcal R$ is the three-dimensional rotation matrix obtained from
\begin{equation}
\mathcal R^{ij}=\tfrac{1}{2}\,\text{tr}_2[D^\dag\sigma^iD\sigma^j].
\end{equation}
These equations show that the standard spinor and four-vector transformation laws are equivalent to a Lorentz transformation of the momentum argument $p\mapsto\Lambda p$ accompanied by a rotation of the canonical polarization argument $\vec s\mapsto\mathcal R\vec s$ (and of any other rest-frame vector if any). Explicit expressions for the Wigner rotations are in general quite complicated, and depend on both the standard boosts and the rest frame, as one can see from Eq.~\eqref{RW}. 

In our approach, all the moving spinors are expressed in terms of the rest-frame ones~\eqref{genericspinor}. Wigner rotation effects should therefore automatically be taken into account. One can then understand that the presence of $0$-components and the auxiliary four-vector $n$ in our expressions is necessary so as to ensure that standard spinor and Lorentz four-vector transformations can alternatively be put in the form of a change of momentum variables accompanied with a rotation of the polarization arguments
\begin{equation}
\begin{split}
S[\Lambda]u(p,\lambda)&=\mathcal N_p\left(\,\uslash p'+m\right)\uslash n'\,S[\Lambda]\chi_\lambda\\
&= \mathcal N_{p'}\left(\,\uslash p'+m\right)\uslash n\sum_{\lambda'}\chi_{\lambda'}\,D_{\lambda'\lambda},
\end{split}
\end{equation}
where the prime on four-vectors indicates that they are Lorentz transformed.
\newline

To summarize, our explicit expressions for the Dirac bilinears~\eqref{S1}-\eqref{T51} are Lorentz covariant since they are directly obtained from explicit spinors transforming in the usual way. The lack of manifest Lorentz covariance stems from the use of canonical polarization which requires a reference to the rest frame. Such a reference is spelled out explicitly in our expresions and appears in the form of $0$-components and auxiliary four-vector $n$. These dependences usually do not appear explicitly because they are contained in the specific form for the spinors $u(p,\lambda)$ and the covariant spin vector $S^\mu(p,\uvec s)$.

\section{Conclusions}\label{sec6}

Using a convenient representation of the massive free Dirac spinors in terms of the rest-frame spinors, we derived new explicit expressions for all the Dirac bilinears. Contrary to most of the expressions proposed in the literature, our results are valid for all momenta and canonical polarizations. In principle, a similar approach can be applied to massless spinors provided one uses a proper representation in term of some reference massless spinors.  

Despite the appearances, our expressions are consistent with Lorentz covariance. The explicit dependence on an external four-vector arises as a consequence of the use of a canonical polarization basis and can be understood in terms of Wigner rotation effects.

The new explicit expressions are particularly well suited for applications in hadronic physics, where Fourier transforms of amplitudes are often needed, requiring a clear distinction between spin and orbital angular momentum. It will also be interesting to see whether these new explicit expressions can improve the efficiency of current codes devoted to the calculation of amplitudes in Particle Physics.

\section*{Acknowledgement}

This work was supported by the Agence Nationale de la Recherche under the project ANR-16-CE31-0019.

\appendix

\section{Explicit expressions}

In this Appendix, we particularize our general results \eqref{S1}-\eqref{T51} to the cases of instant-form and light-front Dirac spinors expressed in the standard (or Dirac) representation.

\subsection{Instant form}

Setting $n_\text{IF}=(1,0,0,0)$ in Eq.~\eqref{genericspinor}, we obtain the canonical Dirac spinors, which read in the standard representation
\begin{align}
u_\text{IF}(p,+)=\frac{1}{\sqrt{E_p+m}}\,\begin{pmatrix}
E_p+m\\
0\\
p_z\\
p_R
\end{pmatrix},\\
u_\text{IF}(p,-)=\frac{1}{\sqrt{E_p+m}}\,\begin{pmatrix}
0\\
E_p+m\\
p_L\\
-p_z
\end{pmatrix}
\end{align}
with $p_{R,L}=p_x\pm ip_y$ and $E_p=p\cdot n_\text{IF}=\sqrt{\uvec p^2+m^2}$. 

We define the average covariant spin-density matrix as
\begin{equation}
\bar S^\mu_\text{IF}=\left(\tfrac{\uvec P\cdot\uvec\sigma}{\bar m},\uvec\sigma+\tfrac{\uvec P\,(\uvec P\cdot\uvec\sigma)}{\bar m(P^0+\bar m)}\right).
\end{equation}
In the forward limit $p'=p$ with equal masses $m'=m$, it reduces to the standard expression~\eqref{SIF}. Introducing the normalization factor
\begin{equation}
\mathcal N_\text{IF}=\tfrac{1}{\sqrt{E_{p'}+m'}\sqrt{E_p+m}}\,\tfrac{m'm}{\bar m^2}
\end{equation}
we find that the Dirac bilinears can be expressed as
\begin{widetext}
\begin{align}
\overline u'_\text{IF}u_\text{IF}&=\left[2(P^2+\bar mP^0)+i\epsilon^{0P\Delta \bar S_\text{IF}}\right]\mathcal N_\text{IF},\\
\overline u'_\text{IF}\gamma_5u_\text{IF}&=(P^0+\bar m)\,(\Delta\cdot \bar S_\text{IF})\,\mathcal N_\text{IF},\\
\overline u'_\text{IF}\gamma^\mu u_\text{IF}&=\left\{2\left[P^\mu(P^0+\bar m)+\tfrac{\Delta^2\eta^{\mu 0}-\Delta^\mu\Delta^0}{4}\right]-i\epsilon^{\mu\tau\Delta \bar S_\text{IF}}(P_\tau+\bar m\delta^0_\tau)\right\}\mathcal N_\text{IF},\\
\overline u'_\text{IF}\gamma^\mu\gamma_5 u_\text{IF}&=\left\{2\left[\bar S^\mu_\text{IF}(P^2+\bar mP^0)-(P^\mu+\bar m\,\tfrac{\Delta^2\eta^{\mu0}-\Delta^\mu\Delta^0}{\Delta^2})\,(P\cdot \bar S_\text{IF})+\tfrac{\Delta^\mu(\Delta\cdot \bar S_\text{IF})}{4}\right]+i\epsilon^{\mu 0P\Delta}\right\}\mathcal N_\text{IF},\\
\overline u'_\text{IF}i\sigma^{\mu\nu}u_\text{IF}&=\left\{2\left[i\epsilon^{\mu\nu\tau \bar S_\text{IF}}(\bar mP_\tau+P^2\delta^0_\tau)-(P_\tau P^{[\mu}-\tfrac{\Delta_\tau\Delta^{[\mu}}{4})\,i\epsilon^{\nu]0\tau \bar S_\text{IF}}+\tfrac{\Delta^0}{\Delta^2}\,i\epsilon^{\mu\nu P\Delta}\,(P\cdot \bar S_\text{IF})\right]\right.\nonumber\\
&\qquad\left.+(P^{[\mu}+\bar m\eta^{0[\mu})\Delta^{\nu]}\right\}\mathcal N_\text{IF},\\
\overline u'_\text{IF}i\sigma^{\mu\nu}\gamma_5u_\text{IF}&=\left\{2\bar S^{[\mu}_\text{IF}\left[P^{\nu]}(P^0+\bar m)+\tfrac{\Delta^2\eta^{\nu]0}-\Delta^{\nu]}\Delta^0}{4}\right]-2\eta^{0[\mu}\left[P^{\nu]}P_\tau-\tfrac{\Delta^{\nu]}\Delta_\tau}{4}\right]\bar S^\tau_\text{IF}-2P^{[\mu}\tfrac{\Delta^{\nu]}\Delta^0}{\Delta^2}\,(P\cdot \bar S_\text{IF})\right.\nonumber\\
&\qquad -i\epsilon^{\mu\nu\tau\Delta}(P_\tau+\bar m\delta^0_\tau)\Big\}\,\mathcal N_\text{IF}.
\end{align}
\end{widetext}

\subsection{Front form}

Setting $n_\text{LF}=(1,0,0,-1)/\alpha$ with $\alpha$ an arbitrary positive constant in Eq.~\eqref{genericspinor}, we obtain the light-front (LF) spinors in the standard representation~\cite{Lepage:1980fj,Brodsky:1997de}

\begin{align}
u_\text{LF}(p,+)=\frac{1}{\sqrt{2\alpha p^+}}\,\begin{pmatrix}
\alpha p^++m\\
p_R\\
\alpha p^+-m\\
p_R
\end{pmatrix},\\
u_\text{LF}(p,-)=\frac{1}{\sqrt{2\alpha p^+}}\,\begin{pmatrix}
-p_L\\
\alpha p^++m\\
p_L\\
-\alpha p^++m
\end{pmatrix}
\end{align}
with $p^+=p\cdot n_\text{LF}$.

We define the average covariant spin-density matrix as
\begin{equation}
\bar S^\mu_\text{LF}=\left[\tfrac{P^+}{\bar m}\,\sigma_z,\tfrac{P^2-\uvec P_\perp^2-2\bar m^2}{2\bar mP^+}\,\sigma_z+\tfrac{\uvec P_\perp\cdot\bar{\uvec S}_\perp}{P^+},\bar{\uvec S}_\perp\right],
\end{equation}
where $\bar{\uvec S}_\perp=\uvec \sigma_\perp+\tfrac{\uvec P_\perp}{\bar m}\,\sigma_z$. Once again, in the forward limit $p'=p$ with equal masses $m'=m$, we recover the standard expression~\eqref{SLF}. Introducing the normalization factor
\begin{equation}
\mathcal N_\text{LF}=\tfrac{1}{\sqrt{p'^+p^+}}\,\tfrac{m'm}{\bar m^2},
\end{equation}
we find that the Dirac bilinears can be expressed as
\begin{widetext}
\begin{align}
\overline u'_\text{LF}u_\text{LF}&=\left[2\bar mP^++i\epsilon^{+P\Delta \bar S_\text{LF}}\right]\mathcal N_\text{LF},\\
\overline u'_\text{LF}\gamma_5u_\text{LF}&=P^+(\Delta\cdot \bar S_\text{LF})\,\mathcal N_\text{LF},\\
\overline u'_\text{LF}\gamma^\mu u_\text{LF}&=\left\{2\left[P^\mu P^++\tfrac{\Delta^2\eta^{\mu +}-\Delta^\mu\Delta^+}{4}\right]-\bar m\,i\epsilon^{\mu+\Delta \bar S_\text{LF}}\right\}\mathcal N_\text{LF},\\
\overline u'_\text{LF}\gamma^\mu\gamma_5 u_\text{LF}&=\left\{2\bar m\left[\bar S^\mu_\text{LF}P^+-\tfrac{\Delta^2\eta^{\mu +}-\Delta^\mu\Delta^+}{\Delta^2}\,(P\cdot \bar S_\text{LF})\right]+i\epsilon^{\mu +P\Delta}\right\}\mathcal N_\text{LF},\\
\overline u'_\text{LF}i\sigma^{\mu\nu}u_\text{LF}&=\left\{2\left[P^2\,i\epsilon^{\mu\nu+ \bar S_\text{LF}}-(P_\tau P^{[\mu}-\tfrac{\Delta_\tau\Delta^{[\mu}}{4})\,i\epsilon^{\nu]+\tau \bar S_\text{LF}}+\tfrac{\Delta^+}{\Delta^2}\,i\epsilon^{\mu\nu P\Delta}\,(P\cdot \bar S_\text{LF})\right]+\bar m\eta^{+[\mu}\Delta^{\nu]}\right\}\mathcal N_\text{LF},\\
\overline u'_\text{LF}i\sigma^{\mu\nu}\gamma_5u_\text{LF}&=\left\{2\bar S^{[\mu}_\text{LF}\left[P^{\nu]}P^++\tfrac{\Delta^2\eta^{\nu]+}-\Delta^{\nu]}\Delta^+}{4}\right]-2\eta^{+[\mu}\left[P^{\nu]}P_\tau-\tfrac{\Delta^{\nu]}\Delta_\tau}{4}\right]\bar S^\tau_\text{LF}-2P^{[\mu}\tfrac{\Delta^{\nu]}\Delta^+}{\Delta^2}\,(P\cdot \bar S_\text{LF})\right.\nonumber\\
&\qquad -\bar m\,i\epsilon^{\mu\nu+\Delta}\Big\}\,\mathcal N_\text{LF}.
\end{align}
\end{widetext}

\end{document}